\documentclass[a4paper]{PoS}
\usepackage{subcaption}
\usepackage[export]{adjustbox}
\usepackage{amsmath}
\usepackage[utf8]{inputenc}

\title{On the two-loop amplitude for $\mathbf{gg\to ZZ}$ production with full top-mass dependence}

\ShortTitle{Top-quark contributions to the $gg\to ZZ$ amplitude at two loops}

\author{\speaker{Bakul Agarwal} and Andreas von Manteuffel\\
        Department of Physics and Astronomy, Michigan State University, East Lansing, MI 48824, USA\\
        E-mail: \email{agarwalb@msu.edu}, \email{vmante@msu.edu}}

\abstract{In this talk, we discuss top-quark contributions to $ZZ$ production through gluon fusion at two loops. We use syzygies and modular arithmetic to compute the reductions to master integrals.
In order to numerically evaluate the amplitude, we express it in terms of finite integrals using a new method to construct integrable linear combinations of divergent integrands.
}

\FullConference{14th International Symposium on Radiative Corrections (RADCOR2019)\\ 
9-13 September 2019\\
		Palais des Papes, Avignon, France}

\begin{document}

\section{Introduction}
$ZZ$ production in gluon fusion is an important process at the LHC.
It is a background to the on-shell production of a Higgs boson with subsequent decay to four leptons through an (off-shell) $Z$ pair.
For off-shell Higgs production, interference effects with continuum diboson production can amount to $O(10\%)$ \cite{Kauer:2012hd}.
Even though the gluon-induced process only starts at one-loop and is formally NNLO for the hadronic process $pp\rightarrow{}ZZ$, its contribution is $O(60\%)$ \cite{Cascioli:2014yka} of the total NNLO correction due to the high gluon luminosity at the LHC.
The NLO corrections to $gg\rightarrow{}ZZ$ were found to be sizable \cite{Caola:2015psa}, resulting in an $O(5\%)$ increase of the total cross section \cite{Grazzini:2018owa}.

The one-loop amplitude for $gg\rightarrow{}ZZ$ has been known for a long time \cite{Glover:1988rg}. The two-loop corrections with massless quarks in the loops were computed in \cite{vonManteuffel:2015msa,Caola:2015ila}.
Top-quark contributions are expected to be relevant, too, in particular at high invariant mass. 
They may therefore also impact indirect Higgs width determinations as proposed in \cite{Caola:2013yja}.
Top-loop contributions to the two-loop amplitudes were calculated in \cite{Melnikov:2015laa,Caola:2016trd} using the heavy top-mass approximation
and in \cite{Campbell:2016ivq} using Pad{\'e} approximants.
In \cite{Grober:2019kuf}, the authors considered the form factors relevant to the interference with the Higgs production amplitude, incorporating also an expansion around the top-quark pair production threshold at two loops.
In this talk, we focus on the calculation of the two-loop amplitude for on-shell $ZZ$ production through gluon fusion with full top mass dependence.

\section{Amplitude and form factors}

\begin{figure}[t]
\begin{subfigure}[c]{0.33\textwidth}
\includegraphics[width=0.7\linewidth, center]{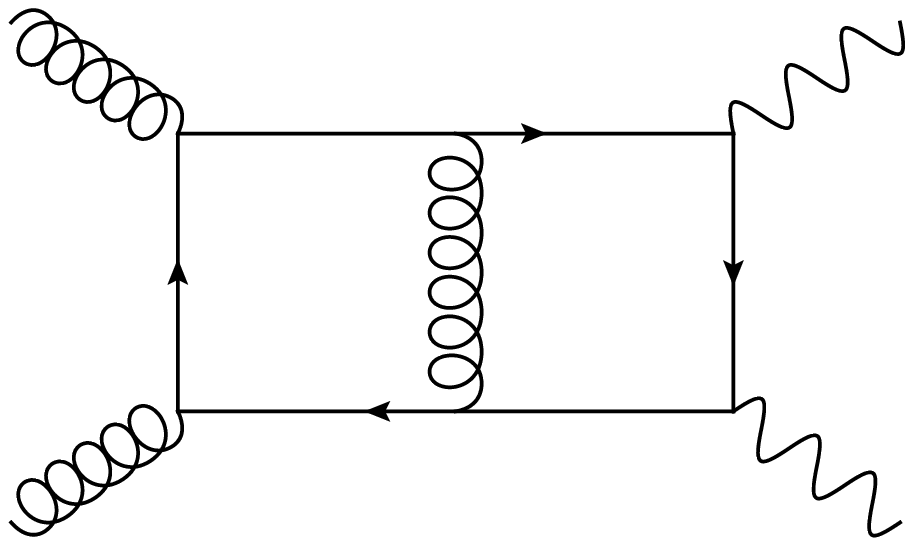}
\caption{}
\label{figure:1:1}
\end{subfigure}
\begin{subfigure}[c]{0.33\textwidth}
\includegraphics[width=0.6\linewidth, center]{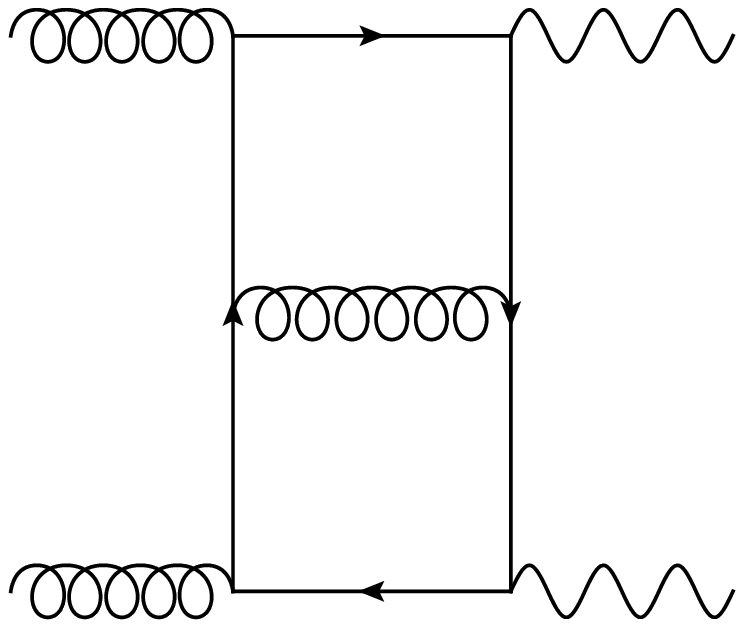}
\caption{}
\label{figure:1:2}
\end{subfigure}
\begin{subfigure}[c]{0.33\textwidth}
\includegraphics[width=0.7\linewidth, center]{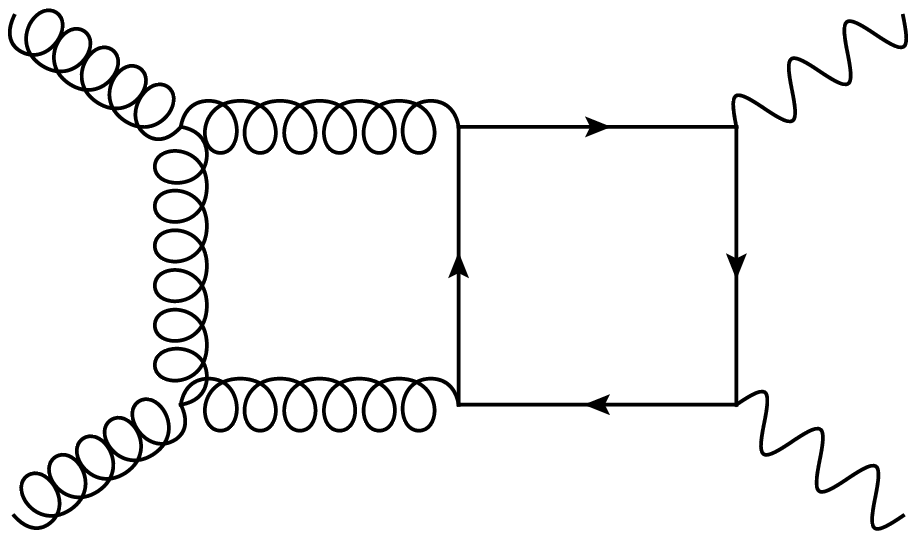}
\caption{}
\label{figure:1:3}
\end{subfigure}
\begin{subfigure}[c]{0.33\textwidth}
\includegraphics[width=0.6\linewidth, center]{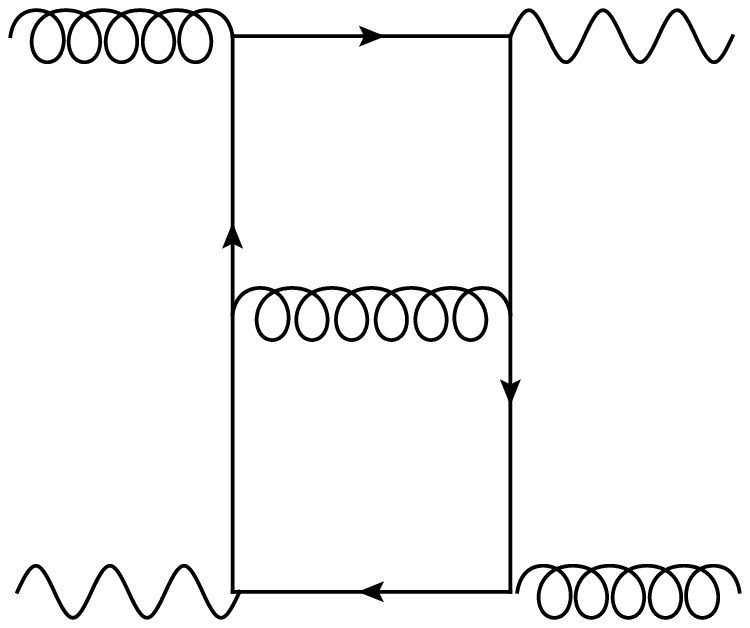}
\caption{}
\label{figure:1:4}
\end{subfigure}
\begin{subfigure}[c]{0.33\textwidth}
\includegraphics[width=0.6\linewidth, center]{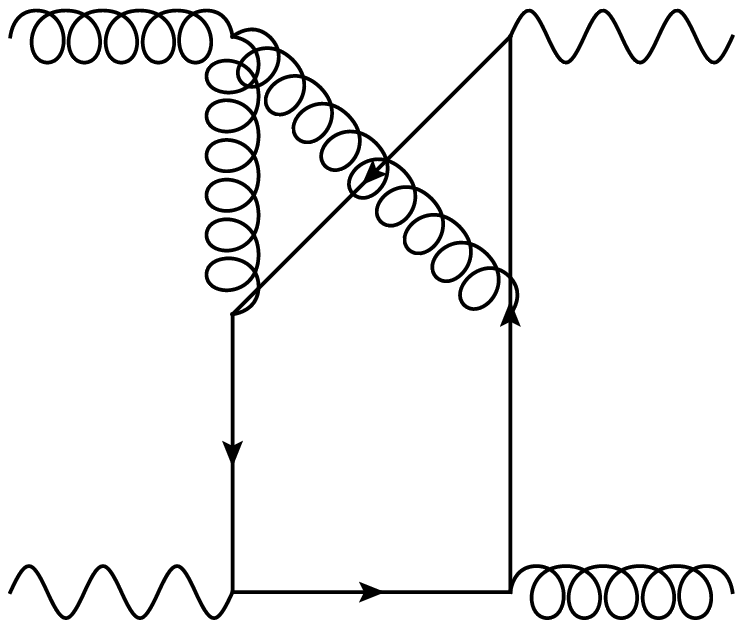}
\caption{}
\label{figure:1:5}
\end{subfigure}
\begin{subfigure}[c]{0.33\textwidth}
\includegraphics[width=0.7\linewidth, center]{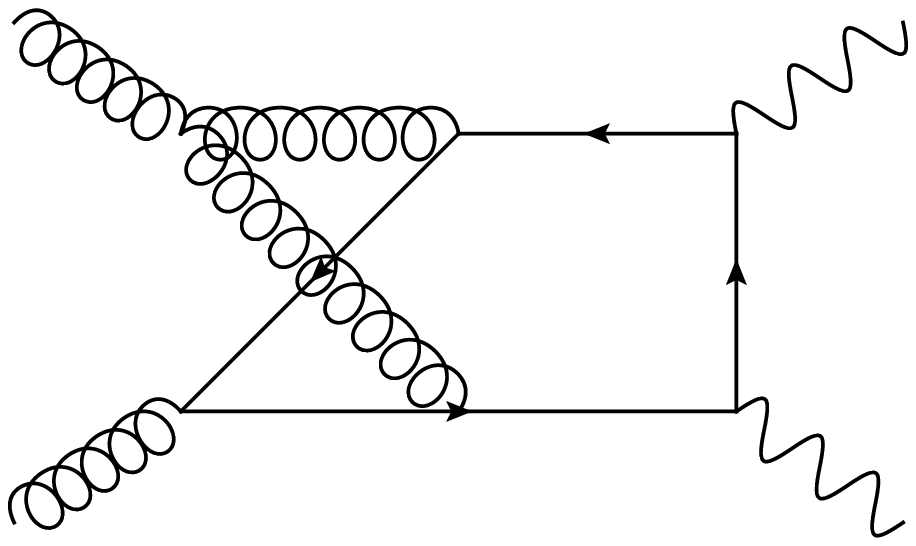}
\caption{}
\label{figure:1:6}
\end{subfigure}
\centering
\begin{subfigure}[c]{0.33\textwidth}
\includegraphics[width=0.7\linewidth, center]{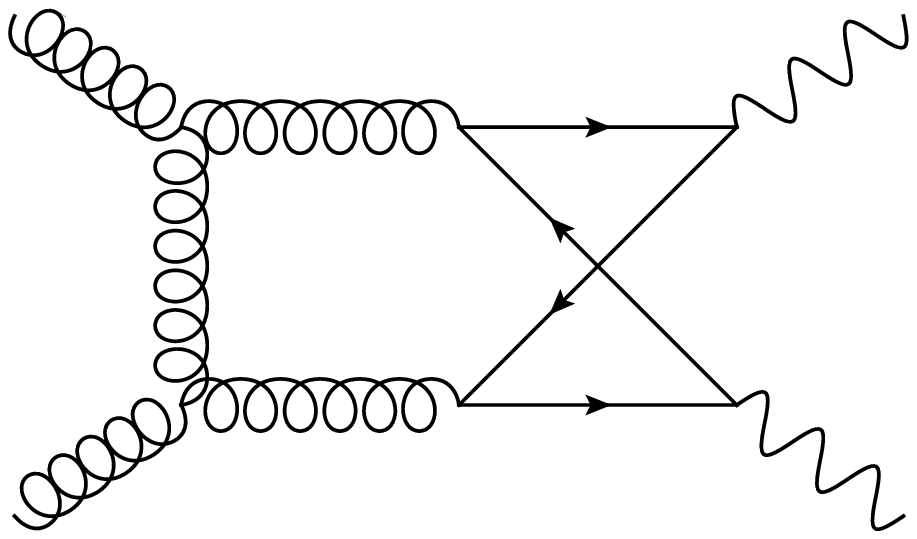}
\caption{}
\label{figure:1:7}
\end{subfigure}
\caption{Representative Feynman diagrams with irreducible topologies.}
\label{figure:1}
\end{figure}

\begin{figure}[t]
\begin{subfigure}[c]{0.33\textwidth}
\includegraphics[width=0.5\linewidth, center]{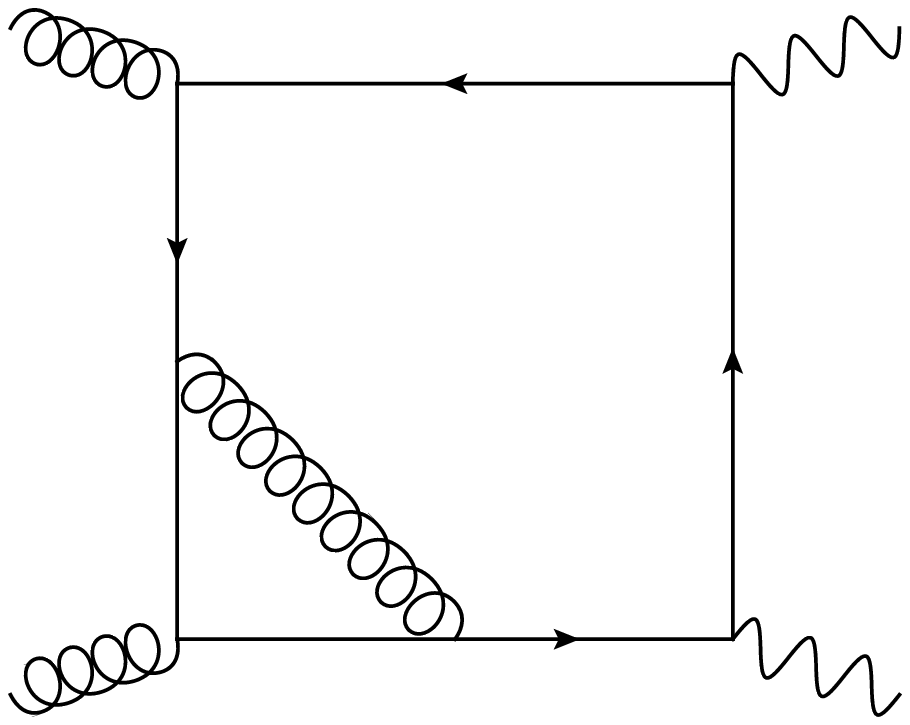}
\caption{}
\label{figure:2:1}
\end{subfigure}
\begin{subfigure}[c]{0.33\textwidth}
\includegraphics[width=0.5\linewidth, center]{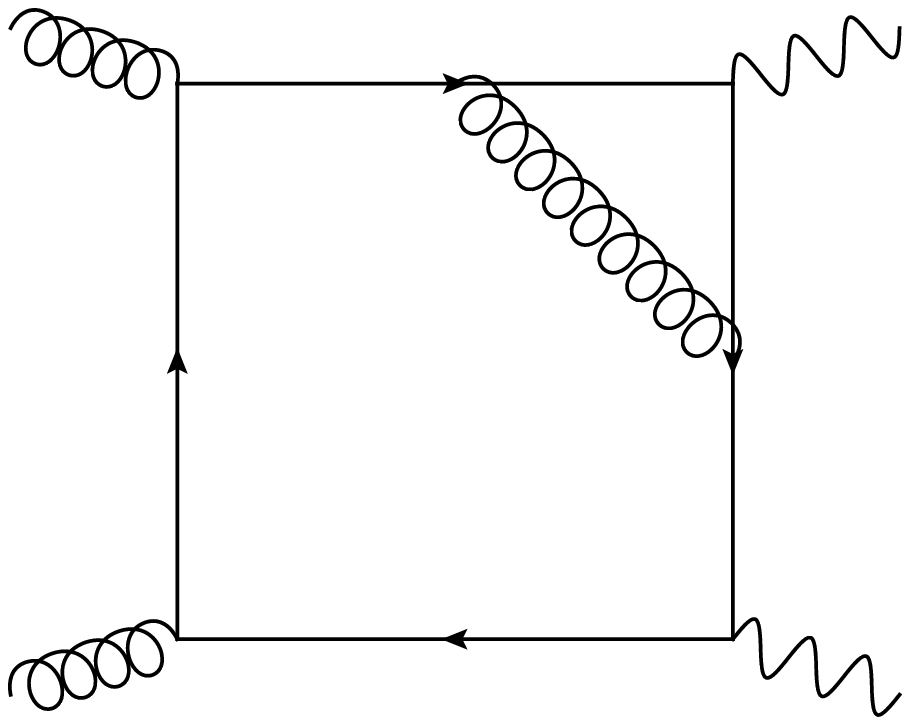}
\caption{}
\label{figure:2:2}
\end{subfigure}
\begin{subfigure}[c]{0.33\textwidth}
\includegraphics[width=0.5\linewidth, center]{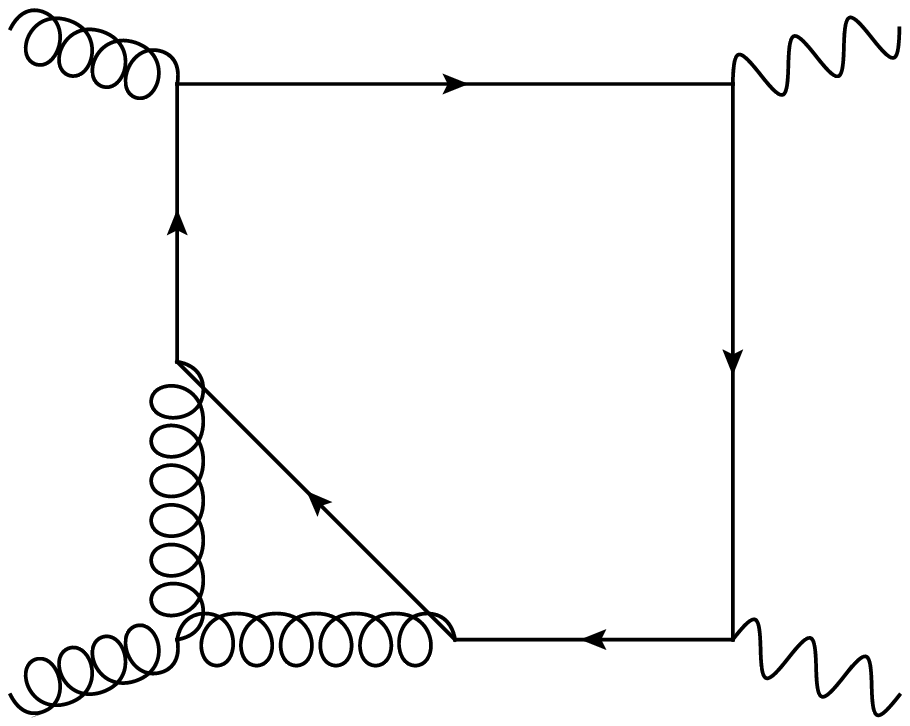}
\caption{}
\label{figure:2:3}
\end{subfigure}
\begin{subfigure}[c]{0.33\textwidth}
\includegraphics[width=0.5\linewidth, center]{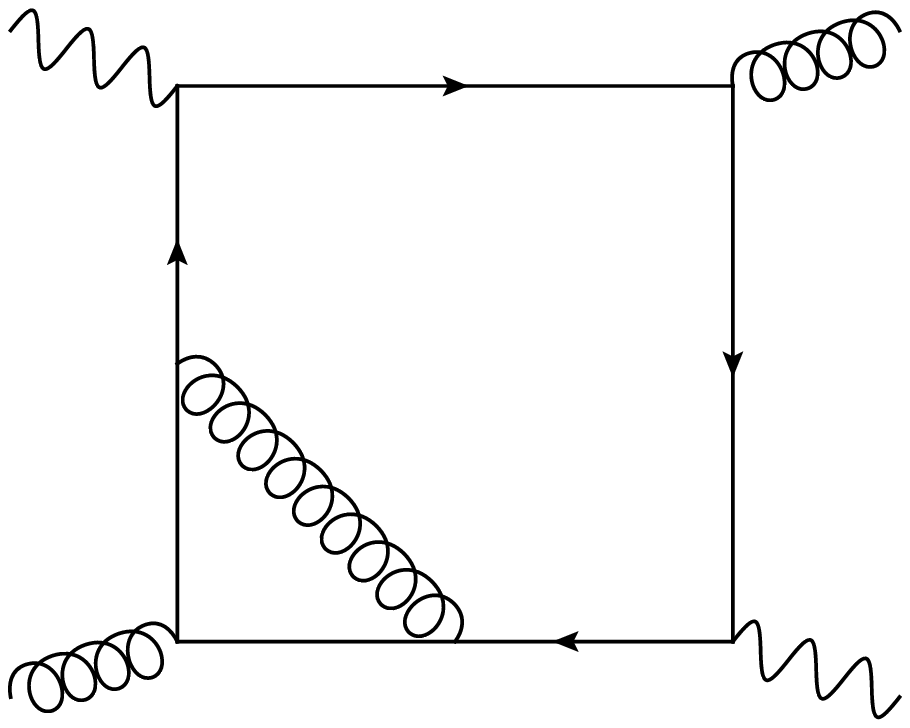}
\caption{}
\label{figure:2:4}
\end{subfigure}
\begin{subfigure}[c]{0.33\textwidth}
\includegraphics[width=0.5\linewidth, center]{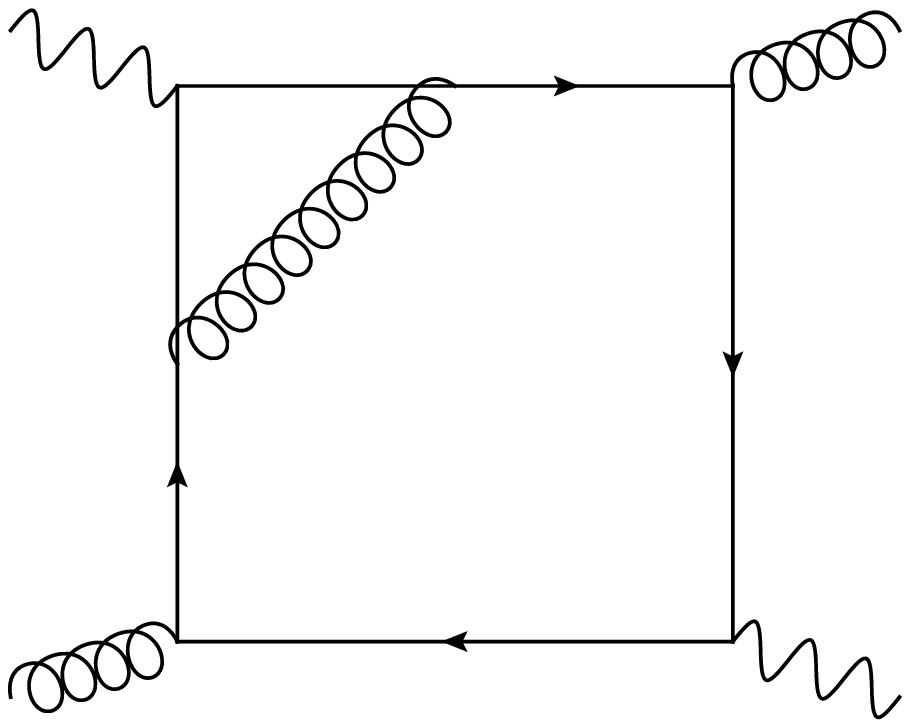}
\caption{}
\label{figure:2:5}
\end{subfigure}
\begin{subfigure}[c]{0.33\textwidth}
\includegraphics[width=0.5\linewidth, center]{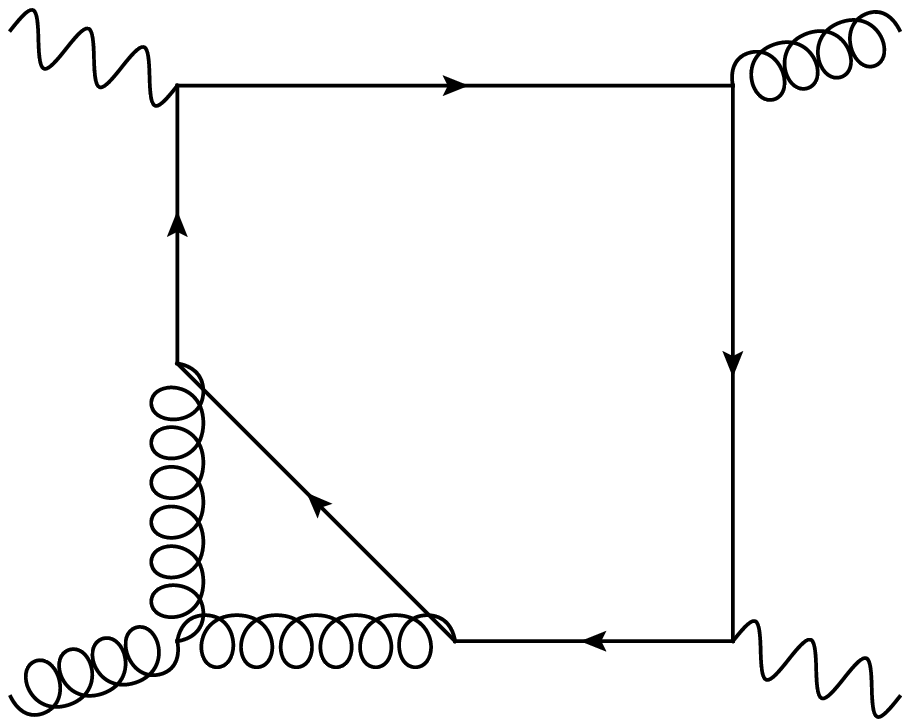}
\caption{}
\label{figure:2:6}
\end{subfigure}
\caption{Representative Feynman diagrams with reducible topologies.}
\label{figure:2}
\end{figure}

We consider the two-loop QCD corrections to the process
\begin{equation}
\label{eq:2:1}
    \qquad\qquad g(p_1) + g(p_2) \rightarrow Z(p_3) + Z(p_4)\,,
\end{equation}
where $p_1$ and $p_2$ are incoming momenta and $p_3$ and $p_4$ are outgoing momenta.
We consider a finite top-quark mass and treat all other quarks as massless.
We employ conventional dimensional regularization  and
decompose the amplitude
\begin{equation}
\label{eq:2:2}
\mathcal{M}=\mathcal{M}_{\mu\nu\rho\sigma}(p_1,p_2,p_3,p_4)\,\epsilon^\mu(p_1)\,\epsilon^\nu(p_2)\,\epsilon^{*\rho}(p_3)\,\epsilon^{*\sigma}(p_4)
\end{equation}\\
into form factors~\cite{vonManteuffel:2015msa} with the unpolarized cross section in mind.
Lorentz invariance allows us to restrict ourselves to 138 tensor structures,
\begin{align}
\label{eq:2:3}
\mathcal{M}^{\mu\nu\rho\sigma}(p_1,p_2,p_3,p_4) &=a_1\,g^{\mu\nu}g^{\rho\sigma}\,+\,a_2\,g^{\mu\rho}g^{\nu\sigma}\,+\,a_3\,g^{\mu\sigma}g^{\nu\rho}\,\notag\\
&+\sum_{i,j=1}^{3}\,(\,a_{1,ij}\,g^{\mu\nu}\,p_i^\rho\,p_j^\sigma\,+\,a_{2,ij}\,g^{\mu\rho}\,p_i^\nu\,p_j^\sigma\,+\,a_{3,ij}\,g^{\mu\sigma}\,p_i^\nu\,p_j^\rho\,\notag\\
&\qquad+\,a_{4,ij}\,g^{\nu\rho}\,p_i^\mu\,p_j^\sigma\,+\,a_{5,ij}\,g^{\nu\sigma}\,p_i^\mu\,p_j^\rho\,+\,a_{6,ij}\,g^{\rho\sigma}\,p_i^\mu\,p_j^\nu\,)\notag\\
&+\sum_{i,j,k,l=1}^{3}\;\;a_{ijkl}\,p_i^\mu\,p_j^\nu\,p_k^\rho\,p_l^\sigma\,.
\end{align}
We can reduce the number of tensors with the following constraints.
We implement transversality of the gluon polarization vectors,
\begin{equation}
\label{eq:2:4}
\epsilon(p_1).p_1 = 0,\qquad\epsilon(p_2).p_2 = 0\,,
\end{equation}
and choose an explicit gauge for the external particles,
\begin{align}
\label{eq:2:5}
\epsilon(p_1).p_2 &= 0, &
\epsilon(p_2).p_1 &= 0, &
\epsilon(p_3).p_3 &= 0, &
\epsilon(p_4).p_4 &= 0\,.
\end{align}
The amplitude can then be written as
\begin{equation}
\label{eq:2:6}
\mathcal{M}^{\mu\nu\rho\sigma}(p_1,p_2,p_3,p_4)\,=\,\sum_{i=1}^{20}\,A_i(s,t,m_t^2,p_3^2,p_4^2)\,T_i^{\mu\nu\rho\sigma}
\end{equation}
with the tensor structures
\begin{align}
T^{\mu\nu\rho\sigma}_1 &= g^{\mu\nu}g^{\rho\sigma} & T^{\mu\nu\rho\sigma}_2 &= g^{\mu\rho}g^{\nu\sigma} & T^{\mu\nu\rho\sigma}_3 &= g^{\mu\sigma}g^{\nu\rho} & T^{\mu\nu\rho\sigma}_4 &= p_1^{\mu}\,p_1^{\nu}\,g^{\rho\sigma}\notag\\
T^{\mu\nu\rho\sigma}_5 &= p_1^{\mu}\,p_2^{\nu}\,g^{\rho\sigma} & T^{\mu\nu\rho\sigma}_6 &= p_2^{\mu}\,p_1^{\nu}\,g^{\rho\sigma} & T^{\mu\nu\rho\sigma}_7 &= p_2^{\mu}\,p_2^{\nu}\,g^{\rho\sigma} & T^{\mu\nu\rho\sigma}_8 &= p_1^{\nu}\,p_3^{\rho}\,g^{\mu\sigma}\notag\\
T^{\mu\nu\rho\sigma}_9 &= p_2^{\nu}\,p_3^{\rho}\,g^{\mu\sigma} & T^{\mu\nu\rho\sigma}_{10} &= p_1^{\mu}\,p_3^{\rho}\,g^{\nu\sigma} & T^{\mu\nu\rho\sigma}_{11} &= p_2^{\mu}\,p_3^{\rho}\,g^{\nu\sigma} & T^{\mu\nu\rho\sigma}_{12} &= p_1^{\nu}\,p_3^{\sigma}\,g^{\mu\rho}\notag\\
T^{\mu\nu\rho\sigma}_{13} &= p_2^{\nu}\,p_3^{\sigma}\,g^{\mu\rho} & T^{\mu\nu\rho\sigma}_{14} &= p_1^{\mu}\,p_3^{\sigma}\,g^{\nu\rho} & T^{\mu\nu\rho\sigma}_{15} &= p_2^{\mu}\,p_3^{\sigma}\,g^{\nu\rho} & T^{\mu\nu\rho\sigma}_{16} &= p_3^{\rho}\,p_3^{\sigma}\,g^{\mu\nu}\notag\\
T^{\mu\nu\rho\sigma}_{17} &= p_1^{\mu}\,p_1^{\nu}\,p_3^{\rho}\,p_3^{\sigma} & 
T^{\mu\nu\rho\sigma}_{18} &= p_1^{\mu}\,p_2^{\nu}\,p_3^{\rho}\,p_3^{\sigma} &
T^{\mu\nu\rho\sigma}_{19} &= p_2^{\mu}\,p_1^{\nu}\,p_3^{\rho}\,p_3^{\sigma} &
T^{\mu\nu\rho\sigma}_{20} &= p_2^{\mu}\,p_2^{\nu}\,p_3^{\rho}\,p_3^{\sigma}\,.
\end{align}
The form factors $A_i(s,t,m_t^2,p_3^2,p_4^2)$ can be related to the amplitude using projection operators $P^{\mu\nu\rho\sigma}_i$ which can themselves be decomposed in terms of the $T_i^{\mu\nu\rho\sigma}$'s as 
\begin{equation}
\label{eq:2:7}
P_i^{\mu\nu\rho\sigma}\,=\,\sum_{j=1}^{20}\,B_{\:ij}(s,t,m_t^2,p_3^2,p_4^2)\,(T_j^{\mu\nu\rho\sigma})^\dagger
\end{equation}
The exact forms of $B_{\:ij}$'s are provided at the \href{https://vvamp.hepforge.org/}{VVamp project website}.
While the above formulae hold in general, in this talk we only consider the top-quark contributions to on-shell $Z$ pair production, i.e.\ $p_3^2=p_4^2=m_Z^2$.
In addition, we use $s=(p_1+p_2)^2$, $t=(p_1-p_3)^2$, and, of course, $p_1^2=p_2^2=0$.
For further simplification, we set the ratio of the Z-boson mass squared and the top-quark mass squared to the numerical value $m_Z^2/m_t^2 = 5/18$.

\begin{figure}[t]
\centering
\begin{subfigure}[c]{0.33\textwidth}
\includegraphics[width=0.4\linewidth, center]{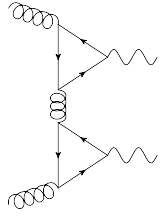}
\label{figure:3:1}
\end{subfigure}
\caption{Representative diagram with triangles generating axial anomaly contributions.}
\label{figure:3}
\end{figure}

We employ {\tt Qgraf} \cite{NOGUEIRA1993279} to generate the Feynman diagrams relevant for the top-quark contributions. In total, we find 174 diagrams out of which 49 vanish due to their colour structure. The remaining diagrams with a single fermion loop can be described using four complete sets of propagators (integral families); there are 13 top-level topologies (trivalent graphs), out of which 7 are irreducible (figure \ref{figure:1}) and 6 are reducible (figure \ref{figure:2}). 

In addition to the diagrams involving a single fermion loop, we also include certain anomalous diagrams involving two closed fermion loops as depicted in \ref{figure:3}. The sum of the vector-vector contribution of these diagrams can be shown to vanish for both massless and massive quarks from Furry's theorem, while the axial-axial piece vanishes only after summing over a degenerate weak SU(2) doublet. For the third generation, the cancellation of the axial-axial part is incomplete and a finite remainder is generated due to the mass splitting for the top-bottom doublet. Therefore, in addition to the top quark contribution, we also need to include the bottom quark contribution to these diagrams for consistency. 

After generating the unreduced amplitude using FORM \cite{Vermaseren:2000nd}, we find 29749 distinct integrals before applying any symmetries, with integrals having up to 5 irreducible scalar products in the numerator. Using {\tt Reduze\;2} \cite{vonManteuffel:2012np,Studerus:2009ye, Bauer:2000cp} we perform a numerical reduction by substituting numbers for kinematic variables and find, after applying all symmetries and sector relations, 286 master integrals, including crossed topologies.

\section{Integration by parts reduction and syzygies}

After performing numerator algebra, we can express the amplitude in terms of scalar integrals.
A general two-loop scalar Feynman integral with N edges can be represented by
\begin{equation}
\label{eq:3:1}
    I(a_1,...,a_N) = \int d^d k_1\,d^d k_2\,\prod^{N}_{i=1}\,\frac{1}{{(q_i^{2}-m_i^{2})}^{a_i}}
\end{equation}
where $k_1, k_2$ are the loop momenta, $q_i$ are the momenta of the edges, $m_i$ are the masses of the edges, $a_i$ are (integer) exponents of the propagators, and $d=4-2\epsilon$. In dimensional regularization, an integral of a total derivative vanishes, which leads to linear relations between integrals with different propagator exponents \cite{Chetyrkin:1981qh}
\begin{equation}
\label{eq:3:2}
    0 = \int d^d k_1\,d^d k_2\,\frac{\partial}{\partial k_j^\mu}\,v^\mu\,\prod^{N}_{i=1}\,\frac{1}{{(q_i^{2}-m_i^{2})}^{a_i}}\,,
\end{equation}
where $v_\mu$ is some linear combination of external and loop momenta. Using these linear relations, we can eliminate most of the integrals appearing in the amplitude with everything expressed in terms of a basis set of integrals, usually referred to as master integrals. An algorithm developed by S.\ Laporta \cite{Laporta:2001dd} can be used to systematically reduce the linear system, and there are many public codes available for this purpose \cite{vonManteuffel:2012np,Smirnov:2014hma,Maierhoefer:2017hyi,Lee:2012cn,Anastasiou:2004vj}.

However, conventional simple choices for the vectors $v_\mu$ lead to the introduction of auxiliary integrals with many additional ``dots'', i.e.\ higher powers $a_i$ of the propagators.
As a consequence, the resulting linear systems are relatively large and computationally expensive to reduce.
A method was proposed in \cite{Gluza:2010ws} to avoid these higher powers of propagators by constructing suitable generating vectors $v_\mu$ from syzygies.
These syzygies may be computed either using a Gr\"obner basis, or up to a specific degree, using linear algebra~\cite{Schabinger:2011dz}.
Subsequent work \cite{Ita:2015tya,Larsen:2015ped} refined these constructions in the momentum space representation but also in Baikov's representation~\cite{Baikov:1996rk}.

Indeed, by a change of variables, the two-loop integral \eqref{eq:3:1} can be written as
\begin{equation}
\label{eq:3:3}
    I(a_1,...,a_N) = \mathcal{N} \int dz_1...dz_N\,\,\frac{1}{\prod^{N}_{i=1}\,z_i^{a_i}}\,\,P^{\frac{d-L-E-1}{2}}\,,
\end{equation}
where $P$ is the Baikov polynomial and $\mathcal{N}$ is a normalization factor,
$L=2$ is the number of loops and $E=3$ is the number of linearly independent external momenta.
In Baikov's representation, integration-by-parts identities are given by
\begin{equation}
\label{eq:3:4}
    0 = \int dz_1...dz_N\,\,\sum^N_{i=1}\,\left(\,\frac{\partial{f_i}}{\partial z_i}\,+\frac{d-L-E-1}{2P}\,f_i\,\frac{\partial{P}}{\partial{z_i}}\,-a_i\,\frac{f_i}{z_i}\right)\,\frac{1}{\prod^{N}_{i=1}\,z_i^{a_i}}\,\,P^{\frac{d-L-E-1}{2}}\,.
\end{equation}
Terms in the above equation that appear with $1/P$ can lead to dimensionally shifted integrals. In order to avoid these integrals, one can impose the constraint
\begin{equation}
\label{eq:3:5}
    \left(\,\sum^N_{i=1}\,f_i\,\frac{\partial{P}}{\partial{z_i}}\,\right)\,+\,g\,P = 0\,.
\end{equation}
This is known as a syzygy constraint in algebraic geometry. Explicit solutions to this equation were pointed out in \cite{Boehm:2017wjc}.
The resulting $f_i$ are simple polynomials of degree 1 in the $z_i$ and the  kinematic invariants.
To enforce the absence of doubled propagators, one requires that for all $i$ with $a_i\ge 1$ the $f_i$ is proportional to $z_i$ to cancel the $1/z_i$ in the relation,
\begin{equation}
\label{eq:3:6}
    f_i\,=\,b_i\,z_i\quad \forall\,i~\text{with~}a_i \geq 1\,.
\end{equation}
While it is straight-forward to fulfil both constraints \eqref{eq:3:5} and \eqref{eq:3:6} separately, a simultaneous solution requires a non-trivial calculation.
Formally, the task at hand is the determination of the intersection of two syzygy modules \cite{Boehm:2018fpv}.
In practice, computer algebra packages implement algorithms to solve this task.
For performance reasons we decided to develop a custom syzygy solver based on linear algebra and finite field arithmetic \cite{vonManteuffel:2014ixa,Peraro:2016wsq}, which converts
the intersection problem up to a specific degree of the syzygies to the row reduction of a matrix.
In this way, we were able to obtain the required syzygies for our calculation.

In a subsequent step, we employed the syzygies to generate linear relations between Feynman integrals with an in-house linear solver based on finite field arithmetic and rational reconstruction.
In this way, we successfully reduced all of the Feynman integrals in our calculation to master integrals.

\section{Finite master integrals using numerators}
After obtaining the reductions to master integrals, the next step is to evaluate them.
A common approach is to evaluate multi-scale master integrals analytically using differential equations \cite{Kotikov:1990kg,Remiddi:1997ny,Gehrmann:1999as,Argeri:2007up,MullerStach:2012mp,Henn:2013pwa,Adams:2017tga}. In this way, the master integrals for the massless two-loop corrections to diboson production were evaluated in \cite{Gehrmann:2014bfa,Gehrmann:2015ora,Henn:2014lfa,Caola:2014lpa}.
It is rather challenging to analytically evaluate the master integrals considered here, however, owing to the presence of the internal mass, as a consequence of which functions beyond multiple polylogarithms are expected.
Instead, we evaluate the master integrals numerically in Feynman parametric representation using Sector Decomposition \cite{Heinrich:2008si,Carter:2010hi}. The conventional basis of master integrals is, however, not adequate for this purpose since these integrals are often divergent and suffer from stability and performance issues in the quadrature. Instead, we choose a different basis composed of integrals which are finite for $d\to 4$~\cite{vonManteuffel:2014qoa,Panzer:2014gra,Bern:1992nf}.
Finite integrals are known to improve numerical integration \cite{Borowka:2016ypz,vonManteuffel:2017myy}. Often, they can be chosen such that fewer terms in the $\epsilon$ expansion of the more complicated topologies are required.

\begin{figure}[t]
\centering
\begin{subfigure}[c]{0.45\textwidth}
\raisebox{-.5\height}{\includegraphics[width=0.6\linewidth, center]{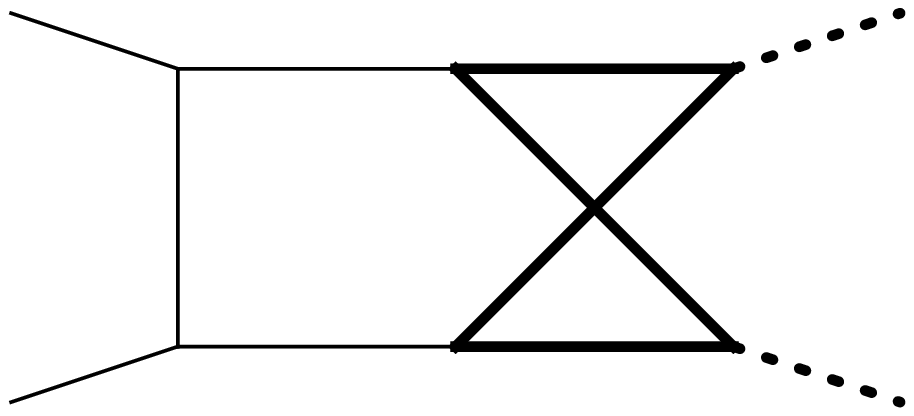}}
\caption{Divergent integral in $d=4-2\epsilon$}
\label{figure:4:1}
\end{subfigure}
\begin{subfigure}[c]{0.45\textwidth}
$(k^2-m_t^2)$\hspace{-1.5cm}\raisebox{-.5\height}{\includegraphics[width=0.6\linewidth, center]{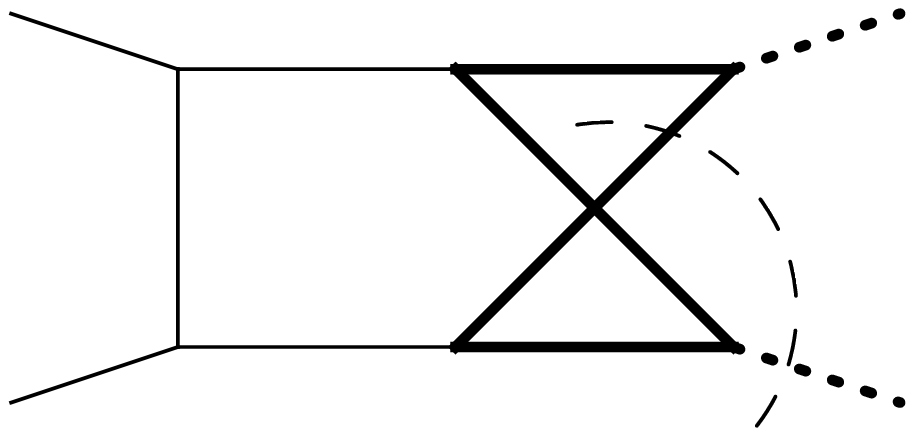}}
\caption{Divergent integral in $d=4-2\epsilon$ with an irreducible numerator}
\label{figure:4:2}
\end{subfigure}
\begin{subfigure}[c]{0.45\textwidth}
\raisebox{-.5\height}{\includegraphics[width=0.6\linewidth, center]{finite_integrals/finD_nonfin.eps}}
\caption{Finite integral in $d=6-2\epsilon$}
\label{figure:4:3}
\end{subfigure}
\begin{subfigure}[c]{0.45\textwidth}
\raisebox{-.5\height}{\includegraphics[width=0.6\linewidth, center]{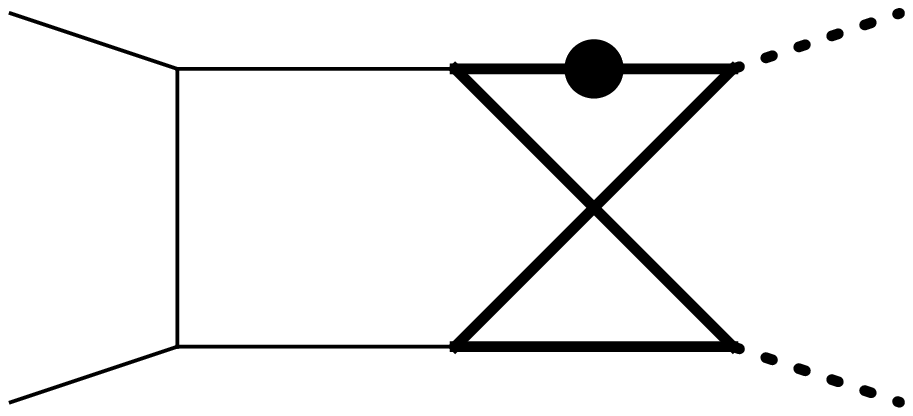}}
\caption{Finite integral with a dot in $d=6-2\epsilon$}
\label{figure:4:4}
\end{subfigure}
\caption{Examples of divergent and finite integrals in the limit $\epsilon \to 0$ for a non-planar topology.
Thick and dashed lines represent massive propagators and legs.
Topology (b) contains an irreducible numerator, where $k$ is the difference of the momenta of the edges marked by the thin dash lines.}
\label{figure:4}
\end{figure}

It is straight-forward to find a basis of finite integrals for any given topology if one allows for dimension shifts and higher powers of existing propagators (dots) e.g.\ by using the finite integral finder implemented in {\tt Reduze\;2}.
Examples for finite integrals of this type are shown in figures \ref{figure:4:3} and \ref{figure:4:4}. In order to change to this basis of integrals one needs to calculate additional reduction identities involving several dots in addition to the usual relations required for the amplitude reduction.
Here, we explore a novel idea of combining divergent integrals to produce finite linear combinations for $\epsilon \to 0$.
The key idea is, that the non-integrable divergences of the different terms cancel at the level of the combined integrand, such that the linear combination can be considered as a single generalized integral for the purpose of numerical integration.
Using our new algorithm we systematically construct such linear combinations, allowing also for integrals with numerators (inverse propagators) and integrals from subtopologies as constituent integrals in $d=4-2\epsilon$ dimensions.
Choosing conventional integrals for the constituent integrals has the advantage that the basis change can be performed using identities very similar to those needed for the regular reduction of the amplitude. 
The numerical performance for such integrals is observed to be on par with those of the finite integrals obtained using dimension shifts and dots, as shown in table \ref{table:2}.

\begin{table}
\centering
\begin{tabular}{ ||c|c|c|| } 
 \hline
 Integral & Relative error & Timing(s) \\ 
 \hline\hline
 Divergent integral in figure \ref{figure:4:1} & $\sim 1$ & 123\\
 \hline
 Divergent integral in figure \ref{figure:4:2} & $\sim 0.5$ & 272\\
 \hline
 Finite integral with dimension shift, in figure \ref{figure:4:3} & $\sim 8\cdot 10^{-4}$ & 81\\
 \hline
 Finite integral with dimension shift and dot, in figure \ref{figure:4:4} & $\sim 2\cdot 10^{-3}$ & 135\\
 \hline
 Finite linear combination in eq. \ref{eq:4:1} & $\sim 2\cdot 10^{-3}$ & 125\\
 \hline
\end{tabular}
\caption{Numerical performance of different non-planar integrals for a physical phase-space point. Timings are for the evaluation of the leading term in $\epsilon$ with {\tt pySecDec} \cite{Borowka:2017idc} using the {\tt VEGAS} algorithm \cite{PETERLEPAGE1978192}.}
\label{table:2}
\end{table}

\begin{figure}[ht]
\centering
\begin{subfigure}[c]{0.45\textwidth}
\raisebox{-.5\height}{\includegraphics[width=0.6\linewidth, center]{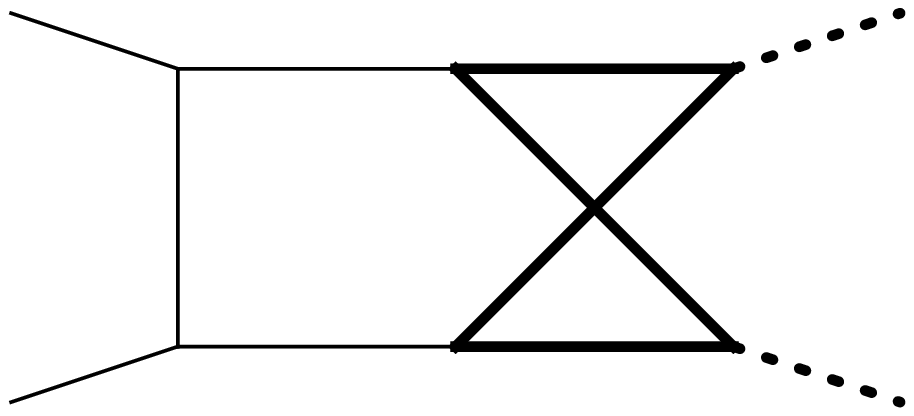}}
\label{figure:5:1}
\caption*{$I_1$}
\end{subfigure}
\begin{subfigure}[c]{0.45\textwidth}
$(k^2-m_t^2)$\hspace{-1.5cm}\raisebox{-.5\height}{\includegraphics[width=0.6\linewidth, center]{finite_integrals/finD_7.eps}}
\label{figure:5:2}
\caption*{$I_2$}
\end{subfigure}
\begin{subfigure}[c]{0.45\textwidth}
\raisebox{-.5\height}{\includegraphics[width=0.6\linewidth, center]{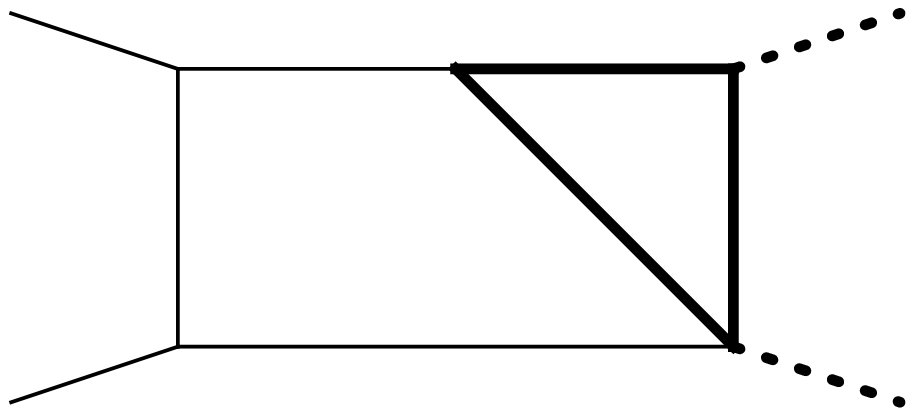}}
\label{figure:5:3}
\caption*{$I_3$}
\end{subfigure}
\begin{subfigure}[c]{0.45\textwidth}
\raisebox{-.5\height}{\includegraphics[width=0.6\linewidth, center]{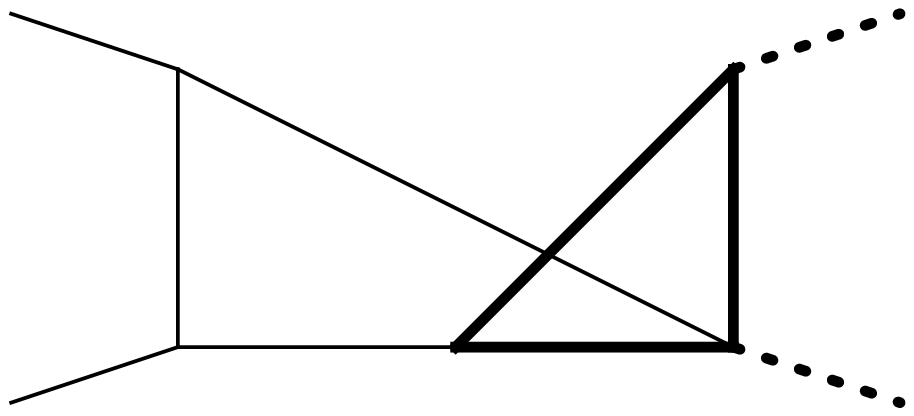}}
\label{figure:5:4}
\caption*{$I_4$}
\end{subfigure}
\begin{subfigure}[c]{0.45\textwidth}
\raisebox{-.5\height}{\includegraphics[width=0.6\linewidth, center]{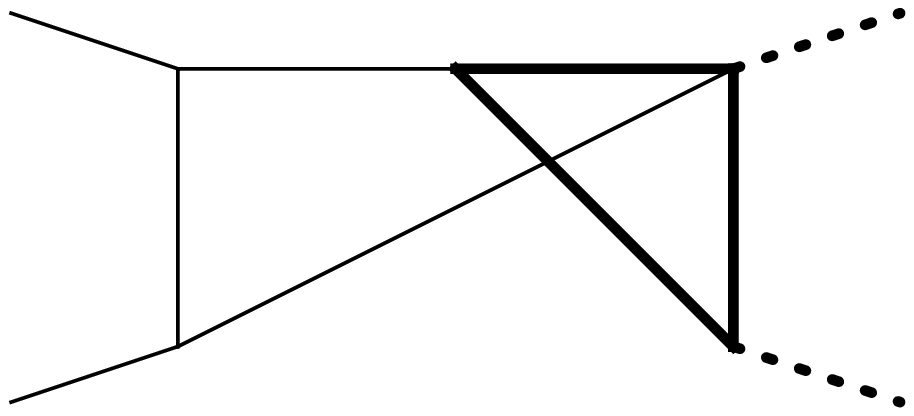}}
\label{figure:5:5}
\caption*{$I_5$}
\end{subfigure}
\begin{subfigure}[c]{0.45\textwidth}
\raisebox{-.5\height}{\includegraphics[width=0.6\linewidth, center]{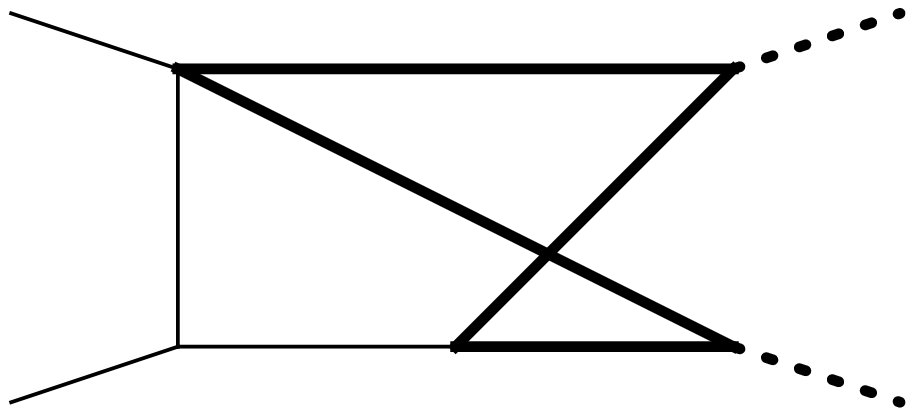}}
\label{figure:5:6}
\caption*{$I_6$}
\end{subfigure}
\begin{subfigure}[c]{0.45\textwidth}
\raisebox{-.5\height}{\includegraphics[width=0.6\linewidth, center]{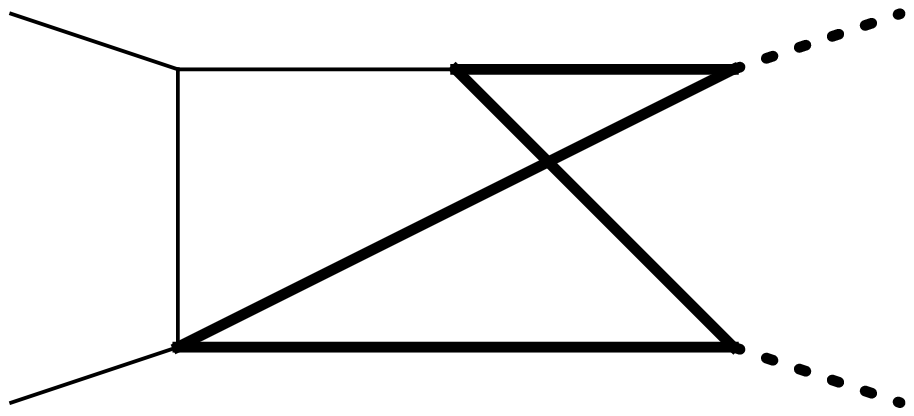}}
\label{figure:5:7}
\caption*{$I_7$}
\end{subfigure}
\caption{Integrals appearing in eq.~\eqref{eq:4:1}.
$I_1$ is the corner integral of the topology under consideration.
$I_2$ is a second integral in the topology, but with a numerator $(k^2-m_t^2)$ where $k$ is equal to the difference of the momenta of the edges marked by the thin dashed lines.
Integrals $I_3,I_4,I_5,I_6,I_7$ belong to subtopologies.
All integrals are defined in $d=4-2\epsilon$ dimensions.}
\label{figure:5}
\end{figure}

As an example, using the constituent integrals shown in figure \ref{figure:5}, we can construct the following finite linear combination:
\begin{equation}
\label{eq:4:1}
    I_{fin}\,\,=\,\,s\,(m_z^2-s-t)\,\,I_1\,+s\,\,I_2\,\,+s\,\,I_3\,-s\,\,I_4\,-s\,\,I_5\,-(m_z^2-s-t)\,\,I_6\,-(m_z^2-t)\,\,I_7\,.
\end{equation}
We find that integrals constructed using such linear combinations of numerator and subtopology integrals are able to cure the (IR) divergences of our more complicated topologies.
For simpler topologies, we also employ finite integrals with dimension shifts and dots, in particular to cure UV poles.
In this way, we express our amplitude in terms of finite master integrals suitable for numerical evaluation.

\section{Conclusions}

Diboson production is an important process for precision measurements at the LHC.
In this talk, we discussed top-quark contributions to onshell ZZ production at two loops with exact dependence on the top-quark mass.
For the reduction of the loop integrals we employed syzygies, which we constructed using linear algebra and modular arithmetic.
We developed a new algorithm to construct a basis of finite integrals through linear combinations of divergent integrands.
The resulting representation of the two-loop amplitude allows for a stable numerical evaluation using {\tt pySecDec} and paves the way for the calculation of precise cross sections and distributions.

\section*{Acknowledgments}
We would like to thank Gudrun Heinrich, Stephan Jahn and Stephen Jones for their help with {\tt pySecDec} and collaborations on related work,
as well as the Max Planck Institute Munich for its hospitality and support.
This work was supported by the National Science Foundation under Grant No.\ 1719863.


\bibliographystyle{JHEP}
\bibliography{refs.bib}

\end{document}